\documentclass[aps,pre,reprint,groupedaddress,preprintnumbers,showpacs,10pt,amsmath,floatfix]{revtex4-1}

\usepackage[T2A,T1]{CJKutf8}

\usepackage[russian,british]{babel}

\usepackage{epsfig}
\usepackage{amssymb}
\usepackage{array}
\usepackage{stmaryrd}
\usepackage{graphicx}

\renewcommand{\phi}{\varphi}
\renewcommand{\theta}{\vartheta}
\renewcommand{\iota}{\ensuremath{\mathbf{\imath}}}
\renewcommand{\epsilon}{\varepsilon}

\newcommand{\abs}[1]{\ensuremath{\left|{#1}\right|}}
\newcommand{\absnum}{\ensuremath{\mathnormal{N}}}
\newcommand{\area}{\ensuremath{\mathnormal{a}}}
\newcommand{\binderchempot}{\ensuremath{\chempot(\equimolar)}}
\newcommand{\block}{\ensuremath{\mathnormal{\lambda}}}
\newcommand{\capillarity}{\ensuremath{\radius_\kappa}}
\newcommand{\charlength}{\ensuremath{\mathnormal{L}}}
\newcommand{\chempot}{\ensuremath{\mathnormal{\mu}}}
\newcommand{\contribution}[1]{\ensuremath{\flat \, {#1}}}
\newcommand{\cutoff}{\ensuremath{\distance_\mathrm{c}}}
\newcommand{\density}{\ensuremath{\mathnormal{\rho}}}
\newcommand{\differential}{\ensuremath{\mathnormal{d}}}
\newcommand{\distance}{\ensuremath{\mathnormal{r}}}
   \newcommand{\vdistance}{\ensuremath{\mathbf{r}}}
\newcommand{\distortion}{\ensuremath{\mathnormal{\xi}}}
\newcommand{\dualtension}{\ensuremath{\mathnormal{\zeta}}}
\newcommand{\eer}{\ensuremath{\mathnormal{\eta}}}
\newcommand{\ellipticity}{\ensuremath{\mathnormal{\Xi}}}
\newcommand{\energy}{\ensuremath{\mathnormal{E}}}
\newcommand{\equimolar}{\ensuremath{\radius_\rho}}
\newcommand{\fed}{\ensuremath{\mathnormal{f}}}
\newcommand{\halfDp}{\ensuremath{\mathnormal{\phi}}}
\newcommand{\helmholtz}{\ensuremath{\mathnormal{A}}}
\newcommand{\inta}{\ensuremath{\mathnormal{i}}}
\newcommand{\intb}{\ensuremath{\mathnormal{j}}}
\newcommand{\kboltz}{\ensuremath{\mathnormal{k}}}
\newcommand{\laplace}{\ensuremath{\radius_\gamma}}
\newcommand{\landau}[1]{\ensuremath{\mathcal{O}\mathnormal{\left({#1}\right)}}}
\newcommand{\liq}[1]{\ensuremath{{#1}'}}
\newcommand{\LJenergy}{\ensuremath{\mathnormal{\epsilon}}}
\newcommand{\LJsize}{\ensuremath{\mathnormal{\sigma}}}
\newcommand{\mardyn}{\textit{mardyn}}
\newcommand{\mass}{\ensuremath{\mathnormal{m}}}
\newcommand{\mindensity}{\ensuremath{\density_\mathrm{min}}}
\newcommand{\parama}{\ensuremath{\mathnormal{\alpha}}}
\newcommand{\paramb}{\ensuremath{\mathnormal{\beta}}}
\newcommand{\partition}{\ensuremath{\mathnormal{Z}}}
\newcommand{\polar}{\ensuremath{\mathnormal{z}}}
   \newcommand{\vpolar}{\ensuremath{\mathbf{z}}}
\newcommand{\potential}{\ensuremath{\mathnormal{u}}}
\newcommand{\pressure}{\ensuremath{\mathnormal{p}}}
   \newcommand{\normal}{\ensuremath{\pressure_\mathrm{n}}}
   \newcommand{\tangential}{\ensuremath{\pressure_\mathrm{t}}}
\newcommand{\probability}{\ensuremath{\mathnormal{\omega}}}
\newcommand{\quantity}{\ensuremath{\mathnormal{q}}}
\newcommand{\radius}{\ensuremath{\mathnormal{R}}}
\newcommand{\sat}[1]{\ensuremath{{#1}_\mathrm{s}}}
\newcommand{\simtime}{\ensuremath{\mathnormal{t}}}
\newcommand{\specenergy}{\ensuremath{f^\mathrm{E}}}
\newcommand{\surface}{\ensuremath{\mathbf{S}}}
\newcommand{\surfaceenergy}{\ensuremath{\mathnormal{\Sigma}}}
\newcommand{\surfacetension}{\ensuremath{\mathnormal{\gamma}}}
   \newcommand{\planartension}{\ensuremath{\surfacetension_0}}
\newcommand{\temperature}{\ensuremath{\mathnormal{T}}}
\newcommand{\tolman}{\ensuremath{\mathnormal{\delta}}}
\newcommand{\vap}[1]{\ensuremath{{#1}''}}
\newcommand{\volume}{\ensuremath{\mathnormal{V}}}

\newcommand{\qq}[1]{{\bf\flqq}#1{\bf\frqq}}
\newcommand{\etal}{\textit{et al.}}

\newcommand{\rus}[1]{\bgroup\fontencoding{T2A}\foreignlanguage{russian}{#1}\egroup}

\newcommand{\LJTS}{LJTS}

\begin{document}
\preprint{Technical report LTD--CME/2011--09/A}
\title{The excess equimolar radius of liquid drops}

\author{Martin Horsch}
\surname{Horsch}
\altaffiliation{Co-affiliated with Imperial College and Universit\"at Paderborn}
\author{Hans Hasse}
\surname{Hasse}
\affiliation{Lehrstuhl f\"ur Thermodynamik, Fachbereich Maschinenbau und Verfahrenstechnik, Technische Universit\"at Kaiserslautern, Erwin-Schr\"odinger-Str.\ 44, 67663 Kaiserslautern, Germany}
\author{Alexander K.\ Shchekin}
\surname{Shchekin}
\affiliation{Department of Statistical Physics, Faculty of Physics, Saint Petersburg State University, ul.\ Ulyanovskaya, Petrodvoretz, 198504 Saint Petersburg, Russia}
\author{Animesh Agarwal}
\surname{Agarwal}
\author{Stefan Eckelsbach}
\surname{Eckelsbach}
\author{Jadran Vrabec}
\surname{Vrabec}
\affiliation{Lehrstuhl f\"ur Thermodynamik und Energietechnik, Institut f\"ur Verfahrenstechnik, Universit\"at Paderborn, Warburger Str.\ 100, 33098 Paderborn, Germany}
\author{Erich A.\ M{\"u}ller}
\surname{M{\"u}ller}
\author{George Jackson}
\surname{Jackson}
\thanks{Corresponding author: G.\ Jackson}
\email{g.jackson@imperial.ac.uk}
\homepage{http://www3.imperial.ac.uk/people/g.jackson}
\affiliation{Department of Chemical Engineering, Centre for Process Systems Engineering, Imperial College London, London SW7 2AZ, United Kingdom}

\date{\today}

\begin{abstract}
   The curvature dependence of the surface tension is related to
   the excess equimolar radius of liquid drops, i.e., the deviation of
   the equimolar radius from that defined with the macroscopic capillarity
   approximation.
   Based on the Tolman [J.\ Chem.\ Phys.\ \textbf{17}, 333 (1949)]
   approach and its interpretation by Nijmeijer \etal{}
   [J.\ Chem.\ Phys.\ \textbf{96}, 565 (1991)], the surface tension of spherical
   interfaces is analysed in terms of the pressure difference due to curvature.
   In the present study, the excess equimolar radius, which can be obtained
   directly from the density profile, is used instead of the Tolman length.
   Liquid drops of the truncated-shifted
   Lennard-Jones fluid are investigated by molecular dynamics simulation in
   the canonical ensemble, with equimolar radii ranging
   from $4$ to $33$ times the Lennard-Jones size parameter $\LJsize$.
   In these simulations, the magnitudes of the excess equimolar radius and the
   Tolman length are shown to be smaller than $\LJsize \slash 2$.
   Other methodical approaches, from which mutually contradicting
   findings have been reported, are critically discussed, outlining possible
   sources of inaccuracy.
\end{abstract}


\maketitle

\section{Introduction}

The macroscopic capillarity approximation consists in neglecting the curvature dependence
of the surface tension $\surfacetension$ of a spherical liquid drop. Accordingly,
the surface tension of a curved interface in equilibrium is approximated by the
value $\planartension$ in the zero-curvature limit,
i.e., for a planar vapour-liquid interface. The Young-Laplace
equation \cite{Young05, Laplace06, RW82} for spherical interfaces
relates the macroscopic surface tension
to a characteristic radius $\capillarity$ of the liquid drop
\begin{equation}
   \frac{\planartension}{\capillarity}
      = \frac{1}{2}\left(\liq\pressure - \vap\pressure\right)
         = \halfDp,
\label{eqn:laplace}
\end{equation}
which will be referred to as the capillarity radius here. Both the factor
$1\slash\capillarity$ and the difference between the liquid
pressure $\liq\pressure$ and the vapour pressure $\vap\pressure$ characterize
the extent by which the surface is curved; the notation
$\halfDp = (\liq\pressure - \vap\pressure)/2$ for half of the pressure difference
is introduced for convenience.
At equilibrium, the temperature is the same for both
phases, and the pressures $\liq\pressure$ and $\vap\pressure$ correspond
to states with the same chemical potential.
The surface tension $\planartension$ of the planar vapour-liquid phase boundary, which is
relatively easy to access experimentally,
couples these two measures of curvature as a proportionality constant.

In combination with an equation of state for the
bulk fluid, microscopic properties such as the radius of a small liquid drop
can thus be deduced from the macro\-scopic state of the surrounding vapour,
i.e., from its supersaturation ratio, and vice versa.
This approach is the most widespread interpretation of the Gibbs theory
of interfaces \cite{Gibbs78a, Gibbs78b}, and it is the point of departure for
the classical nucleation theory (CNT) as introduced by Volmer and Weber
\cite{VW26} and further developed by Farkas \cite{Farkas27} as well as
subsequent authors \cite{BD35, Kuhrt52, FRLP66}.
With the Gibbs approach one presumes a sharp
dividing surface between the phases, a conceptual picture that does not
reflect the physical phenomena present at the molecular length scale. However, this
abstraction is precisely its strength. Instead of discussing
thermodynamic properties such as the density, the pressure tensor and
the free energy density in a localized way, interfacial
excess quantities can be assigned to the formal dividing surface as a
whole. 

It should be recognized that significant size effects on interfacial properties
had already been detected experimentally by Weber \cite{Weber01} at the
turn of the last century. 
This was also known to Farkas \cite{Farkas27} who stated explicitly that
the capil\-larity approximation should be expected to fail for radii at the
length scale of the intermolecular interactions.
In the absence of a better approximation, however, the surface tension of
the planar phase boundary had to be used for nucleation theory, and little
has changed in this respect in the meantime.

In case of significant deviations from the macroscopic capillarity approximation,
liquid drops cannot be characterized sufficiently by a single effective radius. Instead,
the capillarity radius $\capillarity$ is distinct from the equimolar
radius $\equimolar$, which is also known as the Gibbs adsorption radius.
For a single-component system, the latter is defined by the zero
excess density criterion
\begin{eqnarray}
   & & \int_{0}^{\equimolar} \differential\polar \, \polar^2 \left[\density(\polar) - \liq\density(\chempot, \temperature)\right] \nonumber\\
      & + & \int_{\equimolar}^{\infty} \differential\polar \, \polar^2 \left[\density(\polar) - \vap\density(\chempot, \temperature)\right] = 0,
\label{eqn:equimolar}
\end{eqnarray}
i.e., by comparing a step function based on the bulk liquid and vapour number densities
$\liq\density(\chempot, \temperature)$ and $\vap\density(\chempot, \temperature)$
as functions of the chemical potential $\chempot$ and the temperature $\temperature$,
respectively, with the microscopic radial density profile $\density(\polar)$.
By convention, the density $\density$ corresponds to the number of particles per
volume here, rather than their mass, and $\polar$ denotes the distance from the
centre of mass of the liquid drop. In the following discussion,
$\temperature$ is treated as a parameter (instead of a variable), so that
total differentials are to be understood as partial differentials at
constant temperature.

For curved interfaces in equilibrium, the
chemical potential deviates from its saturated value $\sat\chempot$ for a flat interface.
In case of a drop, both phases are supersaturated. To realize this, it is sufficient
to consider the Gibbs-Duhem equation for a curved phase boundary
\begin{equation}
   \differential\left(\liq\pressure - \vap\pressure\right)
      = \left(\liq\density - \vap\density\right) \differential\chempot.
\end{equation}
For a planar interface, both phases coexist at the saturation condition
($\chempot$ = $\sat\chempot$) and the pressure difference
is zero. Raising the value of the liquid pressure $\liq\pressure$
over the vapour pressure $\vap\pressure$ therefore increases
the chemical potential $\chempot$, which must be equal for both phases in (stable
or unstable) equilibrium, so that its value for a system with a liquid drop
will exceed $\sat\chempot$. The precise conditions can
be determined from the pressure difference between the fluid phases by means of an
equation of state. 

Beside $\capillarity$ and $\equimolar$, a thermodynamically relevant definition
of the liquid drop size is given by the surface of tension radius
\begin{equation}
   \laplace = \frac{\surfacetension}{\halfDp},
\label{eqn:laplaceR}
\end{equation}
which is also known as the Laplace radius. It can be obtained
by inserting the actual value of the surface tension $\surfacetension$
of the system with the curved interface (not the planar limit
value) into the Young-Laplace equation. This radius can be related to the surface
area $\area$ and to the volume $\volume$ of the drop
\begin{equation}
   \laplace \, \differential\area = 2 \, \differential\volume.
\end{equation}
The excess grand potential $\surfaceenergy$ of
the surface thus evaluates to
\begin{equation}
   \laplace \, \differential\surfaceenergy = 2 \surfacetension \, \differential\volume,
\label{eqn:surfaceenergy}
\end{equation}
in terms of the surface tension
\begin{equation}
   \surfacetension = \frac{\differential\surfaceenergy}{\differential\area}.
\end{equation}
Modified versions of the Young-Laplace equation, which
allow for the use of different
radii in an analogous way, were introduced by Buff \cite{Buff51, Buff55}
and Kondo \cite{Kondo56}.

The present study deals with the deviation between the capillarity
radius $\capillarity$, the equimolar radius $\equimolar$ and the
surface of tension radius $\laplace$ of a liquid drop in equilibrium
with a supersaturated vapour. As Tolman \cite{Tolman48, Tolman49a, Tolman49b},
following Gibbs, showed on the basis of axiomatic thermodynamics,
one of these differences, now commonly referred to as the Tolman length
\begin{equation}
   \tolman = \equimolar - \laplace,
\label{eqn:deftolman}
\end{equation}
is sufficient to characterize the curvature dependence of the
surface tension \cite{Tolman49b}
\begin{equation}
   \frac{\differential\ln\laplace}{\differential\ln\surfacetension}
      = 1 + \frac{1}{2}\left(\frac{\tolman}{\laplace} + \left[\frac{\tolman}{\laplace}\right]^2
         + \frac{1}{3}\left[\frac{\tolman}{\laplace}\right]^3 \right)^{-1}.
\label{eqn:tolman}
\end{equation}
It is important to point out that this relation is exact,
strictly following the approach of Gibbs, i.e., without neglecting any of the
higher-order curvature terms. The cubic expression derives from an integral over the spherical
density profile.
However, Eq.\ (\ref{eqn:tolman}) is often transformed into a polynomial expansion for
$\planartension\slash\surfacetension$, which contains an infinite number of terms and has to be
truncated, e.g., after the second-order contribution in terms of
curvature \cite{BDOVB10}
\begin{equation}
   \frac{\planartension}{\surfacetension} = 1
      + \frac{2\tolman_0}{\laplace} + 2\left(\frac{\block}{\laplace}\right)^2
         + \landau{\laplace^{-3}}.
\label{eqn:block}
\end{equation}
Here, $\tolman_0$ is the Tolman length in the zero-curvature (infinite radius) limit.
Castellanos \etal\ \cite{CTG09} have conjectured that \qq{the Tolman length
is related to the interfacial width $\Delta^\sigma$ according to
$\Delta^\sigma \approx 2\delta$.}
The Block length $\block$, which characterizes the effect of Gaussian curvature
that becomes predominant when $\tolman$ is very small or for
systems where, due to an inherent symmetry, $\tolman = 0$ holds by construction,
has recently been investigated by Block \etal\ \cite{BDOVB10}; a similar leading term,
proportional to $\laplace^{-2} \ln\laplace$, has also been deduced by Bieker and
Dietrich \cite{BD98} from DFT based on a Barker-Henderson perturbation expansion.

%

%
One should keep in mind that the Tolman equation as given by
Eq.\ (\ref{eqn:tolman}) is valid for curved phase
boundaries of pure fluids in general, whereas
truncated polynomial expansions in terms of the
curvature $1\slash\laplace$ like Eq.\ (\ref{eqn:block})
necessarily break down for liquid drops at the molecular length scale.
In practice, one of the major problems of the Tolman approach is that
it analyses the surface tension in terms of the radii $\equimolar$
and $\laplace$. While $\equimolar$ can be immediately
obtained from the density profile, $\laplace$ is by definition
related to $\surfacetension$ itself. Since for highly curved
interfaces the value of $\surfacetension$ is disputed
or unknown \cite{SVWZB09, SMMMJ10}, the surface of
tension radius $\laplace$ is correspondingly uncertain.

To resolve this issue, we reformulate Tolman's theory
in terms of $\capillarity$ and $\equimolar$. This leads to greater
transparency, since the capillarity radius $\capillarity$ can be obtained
on the basis of the surface tension in the planar limit $\planartension$,
which is experimentally accessible, and properties of the (stable
and metastable) bulk fluid. It is related to the pressure
difference between the coexisting phases in equilibrium, which
is a bulk property as well, since it can be determined from
$\chempot$ and $\temperature$ with an equation of state for the fluid.
All information on the molecular structure of the curved interface can
thus be captured by a single undisputed quantity here,
namely the equimolar radius $\equimolar$.

For this approach, the excess equimolar radius, defined as
\begin{equation}
   \eer = \equimolar - \capillarity,
\label{eqn:defeta}
\end{equation}
plays a role similar to the Tolman length, and the macroscopic
quantity $\halfDp$ is used instead of $1\slash\laplace$ as a measure
of the influence of curvature on the thermophysical properties of the
interface and the bulk phases. In this way, the thermodynamics of
liquid drops are discussed by following a new route that relies on the
density profiles and bulk properties only, avoiding the intricacies
of defining the pressure tensor or the change in the surface area
as required by other approaches.

\begin{figure}[t!]
\centering
\includegraphics[width=8.667cm]{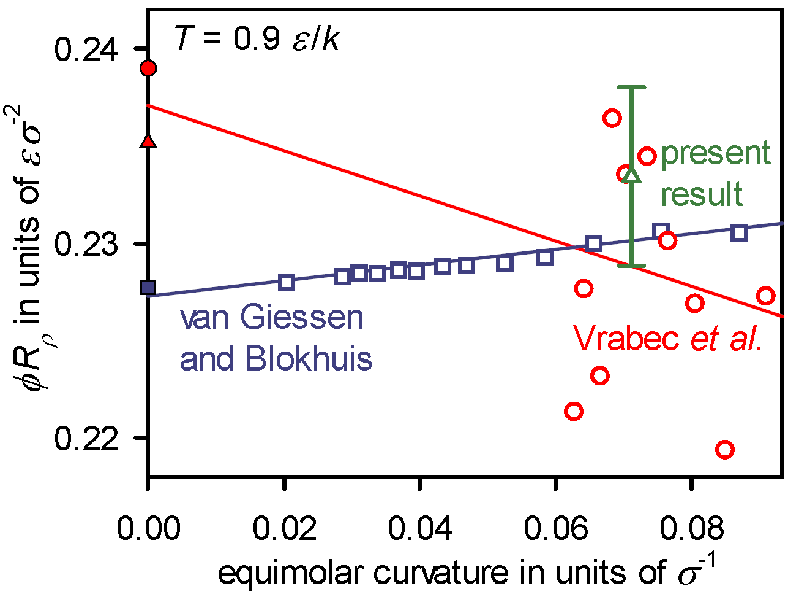}
\caption{
   Representation of van Giessen and Blokhuis \cite{GB09}, showing $\halfDp\equimolar$
   as a function of the equimolar curvature $1\slash\equimolar$ for liquid drops of the
   truncated-shifted Lennard-Jones fluid at $\temperature$ = $0.9$ $\LJenergy\slash\kboltz$,
   where the equimolar radius
   $\equimolar$ is determined from the density profiles and $\halfDp$ from the
   difference between the values of the normal component of the Irving-Kirkwood
   pressure tensor in the homogeneous regions inside the liquid drop as well
   as outside, i.e., in the homogeneous supersaturated vapour. In comparison with
   the results of van Giessen and Blokhuis ($\square$), the data of Vrabec \etal\ \cite{VKFH06}
   ($\circ$), which were obtained by the same method,
   are included here along with a data point ($\triangle$)
   where $\halfDp$ is determined by MD simulation
   of the homogeneous fluid.
   The data for the planar surface tension $\planartension$ are taken from
   simulations of Vrabec \etal\ ($\bullet$) and van Giessen
   and Blokhuis ($\blacksquare$) as well as the
   correlation of Vrabec \etal\ ($\blacktriangle$).
   The continuous lines are guides to the
   eye: In the planar limit, a positive slope corresponds to a negative Tolman
   length and vice versa, cf.\ Eq.\ (\ref{eqn:van-giessen-blokhuis}).
}
\label{fig:VGB}
\end{figure}

The present method is related to the \qq{direct determination} of
$\tolman_0$ proposed by Nijmeijer \etal{} \cite{NBWBL91},
as recently applied by van Giessen and Blokhuis \cite{GB09} on the
basis of a representation of $\halfDp\equimolar$ over
$1\slash\equimolar$ with
\begin{equation}
   - \tolman_0 = \frac{1}{\planartension} \left(
      \lim_{\equimolar\to\infty}
         \frac{\differential}{\differential(1\slash\equimolar)} \,
            \halfDp\equimolar \right),
\label{eqn:van-giessen-blokhuis}
\end{equation}
as depicted in Fig.\ \ref{fig:VGB}.
However, the implementation suggested here is methodologically
different from that of van Giessen and Blokhuis which relies on a
pressure tensor to obtain $\halfDp$, whereas in the present work,
the pressure difference is determined by molecular dynamics (MD)
simulation of the bulk fluids. Applying the
definitions of the capillarity radius and the excess equimolar
radius, Eq.\ (\ref{eqn:van-giessen-blokhuis}) transforms to
\begin{equation}
   - \tolman_0 = \lim_{\equimolar\to\infty} 
      \frac{\differential(\equimolar\slash\capillarity)}{\differential(1\slash\equimolar)}
         = \lim_{\equimolar\to\infty} 
            \frac{\differential(\eer\slash\capillarity)}{\differential(1\slash\equimolar)},
\label{eqn:van-giessen-blokhuis2}
\end{equation}
facilitating an analysis of interface properties in terms of the radii
$\capillarity$ and $\equimolar$ as well as the deviation $\eer$
between them.

This article is structured as follows:
In Section \ref{sec:review}, a review is made of the available routes to the Tolman length
and the surface tension by molecular simulation. MD simulation methods immediately related to
nucleation itself, from which information of the excess free energy of curved interfaces
can also be deduced \cite{TF98, MKEY07, VHH09, HLLTW10}, are not included in
that discussion; in this regard, the reader is referred to
Chkonia \etal\ \cite{CWSWR09}.
Section \ref{sec:eer} is dedicated to a brief outline of how Tolman's thermodynamic approach
can be transformed by analysing the surface tension in terms of $\eer$ and $\halfDp$ rather
than $\tolman$ and $1 \slash \laplace$. The methodology
and the results of a series of canonical ensemble MD
simulations, where the excess equimolar radius is obtained solely on the basis of density
profiles, are presented in Section \ref{sec:method}. An interpretation of
these results is given in Section \ref{sec:discussion}, placing the
present findings in the context of the multitude of
mutually contradicting hypotheses proposed in the literature.


\section{The Tolman length from molecular simulation}
\label{sec:review}

\subsection{Analysis of the planar interface}

For the planar interface, the definition of the Tolman length given by
Eq.\ (\ref{eqn:deftolman}) ceases to be applicable, as the surface
of tension radius $\laplace$ becomes ill-defined in the absence of
curvature, since the pressure is equal on both sides of the interface
in this case.
Therefore, the planar interface Tolman length $\tolman_0$ necessarily
has to be derived from considerations pertaining to curved geometries.
It can be obtained either by extrapolating results for $\tolman$ to
the macroscopic limit $\halfDp \to 0$ (i.e., $\laplace \to \infty$)
or by constructing the limit explicitly from expressions for the
radii $\equimolar$ and $\laplace$. The latter approach was followed
by Fisher and Wortis \cite{FW84} who, on the basis of Landau
(square-gradient) theory, derived the relation
\begin{eqnarray}
   - \tolman_0 & = &
      \left( \int_{\polar = -\infty}^{\polar = \infty} \differential\density_0(\polar)
         \frac{\differential\density_0(\polar)}{\differential\polar} \right)^{-1}
            \int_{\polar = -\infty}^{\polar = \infty} \differential\density_0(\polar)
               \frac{\differential\density_0(\polar)}{\differential\ln\polar} \nonumber\\
   & & + \frac{1}{\Delta\density} \int_{\polar = -\infty}^{\polar = \infty} \differential\density_0(\polar) \, \polar,
\label{eqn:planartolman}
\end{eqnarray}
in terms of the density profile $\density_0(\polar)$ of the planar
interface. This expression can also be
extended to account for the pair density profile, whereby Eq.\
(\ref{eqn:planartolman}) becomes a limiting case \cite{BB93, BS99}.

The available computational methods for evaluating the Tolman length of curved
interfaces, however, involve the determination of the surface
tension $\surfacetension$. It is usually the methodo\-logy related
to the evaluation of $\surfacetension$ that is
both the crucial and the most debatable step, which is made evident
by the contradictory findings for $\surfacetension$ (and
consequently also for $\tolman_0$) obtained from different methods.
Three routes to the surface tension of liquid drops will now be
discussed briefly: the mechanical route as implemented by Thompson
\etal\ \cite{TGWCR84}, the grand canonical route of Schrader \etal\
\cite{SVB09} and the variational route developed by Sampayo \etal\
\cite{SMMMJ10}.

Many different versions and combinations of these approaches
exist \cite{GJBM05, GGLM08, GM11}, but it would be inappropriate to
attempt a full appreciation of the complete body of work here.
The reader is directed to the excellent review by Henderson
\cite{Henderson86} for a detailed discussion of the underlying
statistical mechanical approaches.
%
%

\subsection{The mechanical route}
\label{subsec:virial}

%
The mechanical route to the surface tension is based on the Bakker-Buff equation 
for spherical interfaces \cite{Bakker28, Harasima53, Buff55, TGWCR84}
\begin{equation}
   \surfacetension = \laplace^{-2} \int_{\,\polar = 0}^{\,\polar = \infty} \differential\polar \,
      \polar^2 \left[\normal(\polar) - \tangential(\polar)\right],
\label{eqn:bakker}
\end{equation}
in terms of the normal component $\normal(\polar)$ and the two (equal)
tangential components $\tangential(\polar)$ of the diagonalized pressure
tensor, which is considered as a spherical average, and where the
integration is from the centre of the drop (\polar = 0). With this
relation one expresses the work required for a reversible isothermal deformation
of the system that leads to an infinitesimal
increase of the surface area at constant volume,
which coincides with the associated free energy difference. It is
sufficient to compute either the normal or the tangential pressure
profile, since both are related by \cite{TGWCR84, BB99}
\begin{equation}
   \frac{\differential\normal}{\differential\ln\polar}
      = 2 \left(\tangential - \normal\right).
\label{eqn:normaltangential}
\end{equation}
At mechanical equilibrium, Eq.\ (\ref{eqn:bakker}) can thus be
transformed to \cite{TGWCR84}
\begin{equation}
   2\surfacetension^3 = -\halfDp^2 \int_{\,\polar = 0}^{\,\polar = \infty}
      \differential\normal(\polar) \, \polar^3,
\label{eqn:thompson}
\end{equation}
a term in which $\laplace$ no longer appears. The surface of
tension radius $\laplace$ can be obtained from the Young-Laplace equation
once the surface tension $\surfacetension$ is known. 

The most widespread implementation of this approach in terms of 
intermolecular pair potentials makes use of the
Irving-Kirkwood (IK) \cite{IK50} pressure
tensor, which was first applied to (spherical) interfaces
by Buff \cite{Buff55} and underlies the simulation studies of Vrabec \etal\
\cite{VKFH06} as well as those of van Giessen and Blokhuis \cite{GB09}.
Its normal component is given by \cite{IK50, TGWCR84}
\begin{equation}
   \normal(\polar) = \kboltz\temperature\density(\polar) + \sum_{\{\inta, \intb\} \in \surface}
      -\frac{\differential\potential_{\inta\intb}}{\differential\distance_{\inta\intb}} \,
         \frac{\abs{\vpolar \cdot \vdistance_{\inta\intb}}}{4\pi\polar^3 \, \distance_{\inta\intb}},
\label{eqn:irvingkirkwood}
\end{equation}
wherein $\kboltz$ is the Boltzmann constant and the summation covers the
set $\surface$ containing all sets of particles $\inta$ and $\intb$
that are connected by a line intersecting a sphere of radius $\polar$ around
the centre of mass. The intersection coordinates relative to the centre of mass
of the liquid drop are represented by $\vpolar$ and the
distance between the particles by
$\vdistance_{\inta\intb}$ with $\distance_{\inta\intb} = \abs{\vdistance_{\inta\intb}}$,
while $-\differential\potential_{\inta\intb} \slash \differential\distance_{\inta\intb}$ is
the force acting between the two particles $\inta$ and $\intb$.

Regarding the mechanical route as described here, various issues arise:
\begin{itemize}
   \item{} It is not clear to what extent the spherical average of
      the pressure tensor succeeds in accounting for the free energy contribution
      of capillary waves, i.e., the excited vibrational modi of
      the interface \cite{HL78, Binder82}.
   \item{} Irving and Kirkwood \cite{IK50} originally proposed their
      expression for the special case of \qq{a single component, single
      phase system}. Its derivation relies on truncating an expansion
      in terms of derivatives of the pair density $\density^{(2)}$
      after the first term, thereby
      disregarding the density gradient completely. For a liquid drop,
      this can lead to inaccuracies: \qq{at
      a boundary or interface \dots{} neglecting 
      terms beyond the first may not be justified} \cite{IK50}.
   \item{} By construction, the mechanical route cannot be separated from the
      assumption of a mechanical equilibrium that underlies both the
      basic approach, i.e., Eqs.\ (\ref{eqn:bakker}) to
      (\ref{eqn:normaltangential}), and the derivation of the IK
      pressure tensor, cf.\ Eq.\ (\ref{eqn:irvingkirkwood}). For
      nanoscopic liquid drops, however, configurations deviating
      from the equilibrium shape correspond to a significant fraction of
      the partition function.
   \item{} The non-unique nature of the pressure tensor, which for
      a planar interface does not have a consequence on the computed
      value of the surface tension \cite{WTRH83}, leads to an
      inconsistent description for a curved interface \cite{SH82,
      Henderson86, GM11}. However, the Harasima pressure tensor
      \cite{Harasima53}, where the set $\surface$ is defined
      differently and the tangential pressure profile
      $\tangential(\polar)$ is computed instead of the normal
      component $\normal(\polar)$, has been found to agree rather
      well with the IK tensor \cite{WTRH83, VKFH06, GGLM08}.
\end{itemize}

\subsection{The grand canonical route}

From an analysis of the canonical partition function and its dependence on
the characteristic length $\charlength$ of other\-wise similar systems, Binder
\cite{Binder82} derived very useful scaling laws for
the probability $\probability(\mindensity)$
of relatively small subvolume to have the density $\mindensity$ corresponding
to a maximum of the local free energy, i.e., the least probable local
density between $\liq\density$ and $\vap\density$.
It follows that \qq{the probability of a \textit{homogeneous} state with
order parameter $\mindensity$ decreases exponentially fast with the
volume} while for cases where the corresponding subvolume is
situated within a phase boundary the probability \qq{decreases
exponentially fast with the interface area} \cite{Binder82}. The surface
excess of the grand potential (per unit surface area) can thus be
determined as 
\begin{equation}
   \specenergy = \lim_{\area\to\infty} \frac{\surfaceenergy}{\area}
      = \kboltz\temperature \lim_{\charlength\to\infty}
         \frac{\ln\probability(\mindensity)}{\area(\charlength)},
\label{eqn:specenergy}
\end{equation}
which is related to the surface tension by
$\surfacetension$ $=$ $\differential\surfaceenergy\slash\differential\area$.
Therein, the term
$\area(\charlength)$ describes the dependence of the surface
area on the characteristic length of the system \cite{Binder82}, e.g.,
$\area(\charlength) = 2 \charlength^2$ for a planar slab in a cubic volume
$\volume = \charlength^3$ with standard periodic boundary conditions.

Small subvolumes of a canonical system in the thermodynamic
limit ($\absnum \to \infty$) are equivalent to systems with
constant $\chempot$, $\volume$ and $\temperature$ so that grand
canonical Monte Carlo (GCMC) simulation can equally be applied.
Umbrella sampling may be used to fully sample the
relevant range of values for the order parameter \cite{TV77,
VM04}, corresponding to the number of particles $\absnum$ present
in the grand canonical system. Thereby, a profile is obtained for
the free energy density $\fed(\absnum)$ or $\fed(\density)$, i.e.,
the dependence of the grand potential per volume unit on the order
parameter.

To analyse liquid drops of a certain
size, however, the limit $\area\to\infty$ cannot be applied
since the area $\area$ of the surface of tension is fixed.
Instead, the surface excess term $\specenergy(\equimolar)$
is determined from expressions based on the equimolar radius \cite{SVB09}
\begin{eqnarray}
   \fed(\density) & = & \frac{\liq{\volume}}{\volume}\liq\density(\equimolar)\binderchempot \nonumber\\
                  & + & \frac{\vap{\volume}}{\volume}\vap\density(\equimolar)\binderchempot
                    +   \frac{4\pi\equimolar^2}{\volume}\specenergy(\equimolar).
\label{eqn:schrader}
\end{eqnarray}
$\liq\volume = 4\pi\equimolar^3\slash{}3$ is the volume associated
with the liquid phase here, $\vap\volume = \volume - \liq\volume$ is the
remainder of the volume and $\liq\density(\equimolar)$ as well
as $\vap\density(\equimolar)$ are bulk densities related to
the liquid drop and the surrounding vapour. The chemical potential $\binderchempot$ is
equal for the vapour and liquid regions, but different from both the saturated bulk value
$\sat\chempot$ and the chemical potential $\chempot$ used for the
grand canonical simulation itself.
This formalism has recently been employed by
Schrader \etal\ \cite{SVB09, SVWZB09} as well as Block \etal\ \cite{BDOVB10},
to examine the interfacial properties of drops, bubbles, and
symmetric mixtures in great detail.

The original method of Binder \cite{Binder82} was developed for
planar interfaces. In the case of systems with a spherical geometry,
the following points should be kept in mind:
\begin{itemize}
   \item{} Following the approach of Schrader \etal\ \cite{SVB09},
      the surface tension $\surfacetension$ can be accessed only indirectly, e.g., from
      Eq.\ (\ref{eqn:surfaceenergy}), based on the surface of tension
      radius $\laplace$ which also has to be obtained in a circuitous
      manner. Thereby, care should be
      taken not to confuse $\specenergy$ with $\surfacetension$, or
      $\equimolar$ with $\laplace$.
   \item{} Since the infinite size limit, cf.\ Eq.\ (\ref{eqn:specenergy}),
      does not apply to nanoscopic liquid drops and the systems under consideration
      can be extremely small, it is not generally possible to neglect the
      contribution of homogeneous configurations to $\fed(\density)$ \cite{Binder82}.
   \item{} The assumption that $4\pi\equimolar^2$ is the surface area associated
      with the surface excess for the grand potential of the system, as in Eq.\ (\ref{eqn:schrader}),
      essentially amounts to applying the macroscopic capillarity approximation.
      Such an approach may be justified under certain circumstances,
      but for investigations of the \textit{deviation} from
      capillarity it is of limited use only.
\end{itemize}
Other umbrella sampling based methods \cite{MGTSCNV10, NKSC10}, which will
not be discussed here in detail, are confronted with similar difficulties,
in particular regarding the relation between the surface tension and the
surface excess free energy.

\subsection{The variational route}
\label{subsec:testarea}

The variational route to the surface tension is based on Bennett's \cite{Bennett76} general
considerations of the molecular simulation of free energies and entropic
quantites. In the canonical ensemble, the free energy
difference $\Delta\helmholtz = \helmholtz_1 - \helmholtz_0$ between two states with
equal $\absnum$, $\volume$ and $\temperature$ is given by the quotient
of the respective canonical partition functions $\partition_0$ and
$\partition_1$, which can be evaluated as averages in terms of
internal energy differences \cite{Bennett76}
\begin{eqnarray}
   \exp\left(\frac{\Delta\helmholtz}{\kboltz\temperature}\right)
      & = & \frac{\partition_0}{\partition_1} \nonumber\\
         & = & \frac{ \left< \min\left(1, \exp\left([\energy_1 - \energy_0]\slash[\kboltz\temperature]\right)\right) \right>_1 }{
            \left< \min\left(1, \exp\left([\energy_0 - \energy_1]\slash[\kboltz\temperature]\right)\right) \right>_0 },
\label{eqn:bennett}
\end{eqnarray}
where the index of the angular brackets denotes the system over which
an ensemble average is taken. Bennett proposed the determination of these
energy differences from \qq{separately-generated samples} \cite{Bennett76}
for $\energy_0$ and $\energy_1$. If the two systems differ in the area of
a phase boundary, then the free energy difference can be related to the
surface tension, assuming that all other deviations between the two states
are accurately taken into account. 

Gloor \etal\ \cite{GJBM05} introduced a version of this approach
where differences between the two states are obtained from a single
simulation run for an unperturbed system with the partition
function $\partition_0$. Corresponding configurations of the second,
perturbed system are generated by performing small affine transformations,
keeping the volume and the number of particles
in both phases constant.
In the limit of an infinitesimal distortion of the system,
Eq.\ (\ref{eqn:bennett}) can be simplified as \cite{Zwanzig54, GJBM05}
\begin{equation} 
   \frac{\Delta\helmholtz}{\kboltz\temperature}
      = -\ln \left< \, \exp\left(\frac{-\Delta\energy}{\kboltz\temperature}\right) \right>_0,
\end{equation}
where $\Delta\energy = \energy_1 - \energy_0$,
as the probability distribution functions of the ensembles corresonding
to the unperturbed and the perturbed system converge,
so that a separate sampling is no longer required.
A third-order expansion in the inverse temperature \cite{Zwanzig54}
\begin{eqnarray}
   \frac{\Delta\helmholtz}{\kboltz\temperature}
      & = & \frac{\left<\Delta\energy\right>}{\kboltz\temperature}
         - \frac{\left<\Delta\energy^2\right> - \left<\Delta\energy\right>^2}{2(\kboltz\temperature)^2} \nonumber\\
            & & + \frac{\left<\Delta\energy^3\right>
               - 3\left<\Delta\energy^2\right>\left<\Delta\energy\right>
                  + 2\left<\Delta\energy\right>^3}{6(\kboltz\temperature)^3},
\label{eqn:zwanzig}
\end{eqnarray}
can be used to increase the
precision of the simulation results \cite{GJBM05, SMMMJ10}.
The surface tension is then immediately obtained from
$\Delta\helmholtz\slash\Delta\area$, since the distortion
of the interface itself (as opposed to its increase in area)
makes a negligible contribution to the free energy
difference \cite{Tolman48}.

In analogy with the Widom test-particle method \cite{Widom63},
this implementation of the variational route is also called
the test-area method \cite{GJBM05, BMMJ08}. Following Sampayo
\etal\ \cite{SMMMJ10}, it can be applied to curved interfaces,
where the affine transformation scales one of the cartesian
axes by the factor $1\slash(1 + \distortion)$ and the remaining ones
by $(1 + \distortion)^{1\slash{}2}$. For $\distortion > 0$,
this creates an oblate shape and the area of the surface
of tension is increased by \cite{Sampayo10}
\begin{equation}
   \frac{\Delta\area}{\pi\equimolar^2} = 
      2(1 + \distortion)
         + \frac{\ln([1 + \ellipticity] \slash [1 - \ellipticity])}{(1 + \distortion)^2\ellipticity}
            + \landau{\frac{\tolman \, \Delta\area}{\equimolar^3}},
\label{eqn:oblate}
\end{equation}
with the ellipticity of the average equimolar surface in the
perturbed system given by
$\ellipticity = [1 - (1 + \distortion)^{-3}]^{1\slash{}2}$.
In the prolate case ($\distortion < 0$), the corresponding term
is $\ellipticity = [1 - (1 - \distortion)^{-3}]^{1\slash{}2}$ with \cite{Sampayo10}
\begin{equation}
   \frac{\Delta\area}{\pi\equimolar^2} = 
      2 \left(\frac{\arcsin\ellipticity}{\ellipticity(1 - \distortion)^{1\slash{}2}}
         - \distortion - 1\right) 
            + \landau{\frac{\tolman \, \Delta\area}{\equimolar^{3}}}.
\label{eqn:prolate}
\end{equation}
It can be shown that the first-order term in Eq.\ (\ref{eqn:zwanzig})
is equivalent to the Kirkwood-Buff \cite{KB49} mechanical route expression
for the surface tension \cite{LH77}. The higher-order terms therefore
presumably capture the deviation between the mechanical and variational routes
due to fluctuations or, equivalently, the contribution of non-equilibrium
configurations to $\surfacetension$. Thus, the higher-order contribution
to Eq.\ (\ref{eqn:zwanzig}) may be related to the closed expression
derived by Percus \etal\ \cite{PPG95} for the deviation between the actual
free energy and an approximation based on the local pressure.

From this point of view, the
following aspects of the method merit further con\-sideration:
\begin{itemize}
   \item{}
      While finite differences of higher order are taken into account for
      the energy, no such terms are considered for the surface area here.
      Clearly, the variance of $\Delta\energy$ is partly
      caused by the variance of $\Delta\area$.
      The use of $\equimolar$ for defining the surface
      area, cf.\ Eqs.\ (\ref{eqn:oblate}) and (\ref{eqn:prolate}),
      may lead to further deviations.
   \item{} The variance of $\Delta\energy$ accounts for
      surface oscillations such as long wave-length
      capillary waves, which directly relate to
      equilibrium properties of the interface and therefore do not depend
      on the statistical mechanical ensemble \cite{Henderson86}.
      However, it can also be influenced by fluctuations regarding
      $\liq\density$ (at constant $\liq\volume$) or 
      $\liq\volume$ (at constant $\liq\density$).
      These modi are ensemble dependent, 
      since they are coupled to the density of the vapour phase.
      Canonically, their amplitude increases with the total volume
      and is ill-defined in the thermodynamic limit $\volume\to\infty$.
      Therefore, the surface tension
      from the variational route may depend on the constraints
      imposed on the system by the ensemble.
   \item{} Although the volume associated with each of the phases
      is invariant for test-area transformations, there is still
      a distortion of the sample with respect to the equilibrium conformation.
      The method is therefore limited to isotropic phases,
      since shearing an anisotropic phase will induce an elastic contribution
      to $\Delta\helmholtz$ from the bulk region as well.
\end{itemize}

%
%
%
%
%

\section{Deviation of the equimolar radius from capillarity}
\label{sec:eer}

From the Tolman equation in its approximate poly\-nomial form, cf.\ Eq.\
(\ref{eqn:block}), the excess equimolar radius $\eer$ can be related to the
Tolman length $\tolman$ by
\begin{eqnarray}
 \eer & = & (\tolman + \laplace) - \capillarity \nonumber\\
      & = & \tolman + \laplace\left(1 - \left[1 + \frac{2\tolman_0}{\laplace} + \landau{\laplace^{-2}}\right]\right) \nonumber\\
      & = & - \tolman + \landau{\laplace^{-1}},
\end{eqnarray}
so that its magnitude in the zero-curvature limit is obtained as
\begin{equation}
   \eer_0 = -\tolman_0,
\end{equation}
which is essentially equivalent to Eq.\ (\ref{eqn:van-giessen-blokhuis2}).

Both in the planar limit and in the presence of curvature effects,
it is therefore possible to express the Tolman relations in terms
of the easily accessible quantities $\eer$ and $\halfDp$, rather
than $\tolman$ and $1\slash\laplace$.
The point of departure for such an expression
is the exact closed form of the Tolman equation, cf.\ Eq.\ (\ref{eqn:tolman}).
It should be recalled that this expression is derived from the Gibbs-Duhem equation,
the Young-Laplace equation and the Gibbs adsorption equation \cite{Tolman49b};
hence, it is based entirely on an axiomatic thermo\-dynamic treatment. As opposed to
truncated power series of the form of Eq.\ (\ref{eqn:block}), the alternative description remains
valid when the radius $\laplace$ becomes similar or smaller in magnitude
than the Tolman length. Polynomial expansions in terms of
$\tolman\slash\laplace$ necessarily fail to capture this limit.

%
\begin{figure}[h!]
\centering
\includegraphics[width=8.667cm]{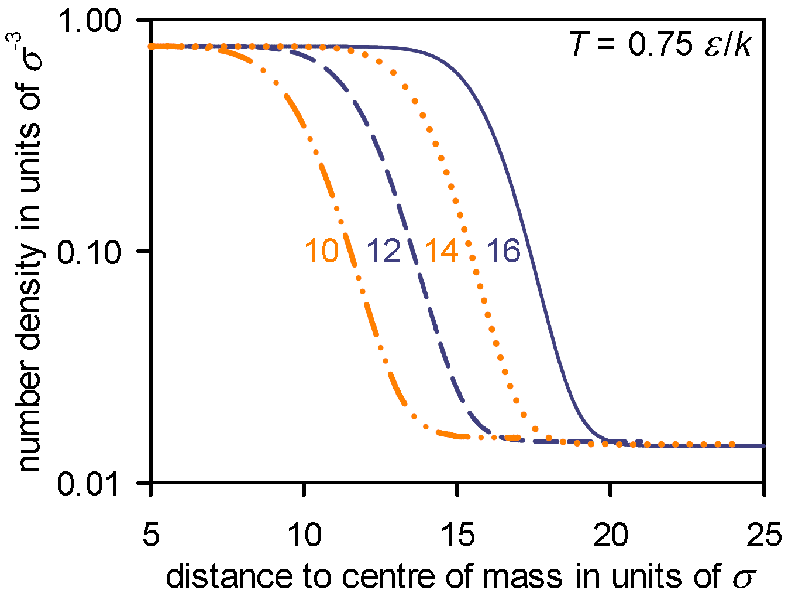}
\caption{
   Density profiles from canonical MD simulations of \LJTS{} liquid
   drops at $\temperature$ = $0.75$ $\LJenergy\slash\kboltz$ with equimolar
   radii of $\equimolar$ = $9.977$ $\pm$ $0.001$ ($\cdot$ --- $\cdot$),
   $12.029$ $\pm$ $0.003$ (-- --), $13.974$ $\pm$ $0.002$ $\LJsize$ ($\cdots$)
   and $15.967$ $\pm$ $0.001$ $\LJsize$ (---), cf.\ Tabs.\
   \ref{tab:error} and \ref{tab:presentsim}.
}
\label{fig:profiles075}
\end{figure}

From the Young-Laplace equation, it follows that
\begin{equation}
   \frac{\differential\laplace}{\differential\halfDp}
      = \frac{1}{\halfDp} \frac{\differential\surfacetension}{\differential\halfDp}
         - \frac{\surfacetension}{\halfDp^2},
\end{equation}
while the reduced length scale appearing in the Tolman equation can be transformed to
\begin{equation}
   \frac{\tolman}{\laplace} = \frac{\eer\halfDp + \planartension}{\surfacetension} - 1,
\end{equation}
by using Eqs.\ (\ref{eqn:laplace}), (\ref{eqn:laplaceR}),
(\ref{eqn:deftolman}) and (\ref{eqn:defeta}). The Tolman relation can thus
be converted to
\begin{eqnarray}
   \frac{\differential\surfacetension}{\differential\halfDp}
      & = & - \frac{2\surfacetension}{\halfDp} \left(
         \frac{\tolman}{\laplace} + \left[\frac{\tolman}{\laplace}\right]^2
            + \frac{1}{3}\left[\frac{\tolman}{\laplace}\right]^3 \right) \\
               & = & \frac{2\surfacetension}{3\halfDp} \left(
                  1 - \left[\frac{\eer\halfDp + \planartension}{\surfacetension}\right]^3
                     \right).
\label{eqn:tolmanetafull}
\end{eqnarray}
This representation of the Tolman result is fully equivalent to Eq.\ (\ref{eqn:tolman}).

For $\halfDp \to 0$, further considerations are required. There,
the curvature dependence of $\surfacetension$ as specified by Eq.\
(\ref{eqn:tolmanetafull}) is only self-consistent under an additional
condition. To demonstrate this, it is helpful to consider
the exact Tolman equation in a different form
\begin{equation}
   \frac{\differential\surfacetension}{\differential\halfDp}
      = \frac{2}{\surfacetension^2} \left(
         \frac{1}{3}\left[\dualtension - \eer^3\halfDp^2\right]
            - \planartension\eer[\planartension + \eer\halfDp] \right),
\label{eqn:tolmandigamma}
\end{equation}
which follows from Eq.\ (\ref{eqn:tolmanetafull}) by
expanding the cubic term where $\dualtension$ has been defined as
\begin{equation}
   \dualtension = \frac{\surfacetension^3 - \planartension^3}{\halfDp}.
\end{equation}
For the sake of brevity, the notation
$\quantity_\inta = \lim_{\halfDp \to 0} \differential^\inta\quantity\slash\differential\halfDp^\inta$
is used here for the $\inta$-th derivative of a quantity $\quantity$
in the zero-curvature limit.
The slope of $\surfacetension$ can be obtained by inserting
\begin{equation}
   \dualtension_0 = \left(\surfacetension^3\right)_1
         = 3\planartension^2\surfacetension_1,
\label{eqn:digamma0}
\end{equation}
into Eq.\ (\ref{eqn:tolmandigamma}), which yields
\begin{equation}
   \surfacetension_1 = 2 \eer_0.
\label{eqn:gammaprime}
\end{equation}
Expanding the excess equimolar radius as
\begin{equation}
   \eer = \eer_0 + \eer_1\halfDp + \landau{\halfDp^2},
\end{equation}
and inserting this expression as well as Eq.\ (\ref{eqn:gammaprime}) into
the planar limit for Eq.\ (\ref{eqn:tolmandigamma}) leads to
\begin{equation}
   \left(\surfacetension^3\right)_2 = 12 \planartension \eer_0^2,
\label{eqn:tolmandigamma2}
\end{equation}
and
\begin{equation}
   \surfacetension_0\surfacetension_2 = -4 \eer_0^2.
\end{equation}
It is by considering
the zero-curvature limit for the
third derivative of $\surfacetension^3$ that a
theo\-rem for the slope of $\eer$ can now be deduced.
Based on Eqs.\ (\ref{eqn:digamma0}) and (\ref{eqn:tolmandigamma2}), a Taylor expansion for
$\differential(\surfacetension^3)\slash\differential\halfDp$ in terms
of $\halfDp$
\begin{equation}
   \frac{\differential}{\differential\halfDp}\surfacetension^3
      = \left(\surfacetension^3\right)_1 + \left(\surfacetension^3\right)_2\halfDp
         + \frac{1}{2}\left(\surfacetension^3\right)_3\halfDp^2 + \landau{\halfDp^3},
\label{eqn:gpower}
\end{equation}
yields
\begin{eqnarray}
   \dualtension & = & \frac{1}{\halfDp} \int_{0}^\halfDp \differential\halfDp \,
      \left(\frac{\differential}{\differential\halfDp} \surfacetension^3\right) \nonumber\\
         & = & 6\planartension^2\eer_0 + 6\planartension\eer_0^2\halfDp
            + \frac{\halfDp^2}{6}\left(\surfacetension^3\right)_3 + \landau{\halfDp^3}.
\label{eqn:taylordigamma}
\end{eqnarray}
From Eqs.\ (\ref{eqn:gammaprime}) to (\ref{eqn:taylordigamma})
\begin{equation}
   6\planartension\left(\planartension\eer_1 + \eer_0^2\right)
      + \frac{\halfDp}{6}\left(\surfacetension^3\right)_3
         = 0 + \landau{\halfDp},
\end{equation}
follows by applying the full Tolman equation, cf.\ Eq.\
(\ref{eqn:tolmandigamma}), in the planar limit. However, this implies
\begin{equation}
   \eer_1 = -\frac{\eer_0^2}{\planartension},
\label{eqn:tolmansuppl}
\end{equation}
which constitutes a necessary boundary condition for
the Tolman approach in terms of $\eer$ and $\halfDp$.

Thus, while there is a direct correspondence between $\tolman_0$ and $\eer_0$,
no such relation exists in case of $\tolman_1$ and $\eer_1$, i.e., the
respective derivatives (in terms of $\halfDp$) in the zero-curvature
limit; instead, $\eer_1$ is fully determined by $\eer_0$ and thus
by $\tolman_0$, the Tolman length of the planar interface. This means
that data on the excess equimolar radius for large radii have a double
significance regarding the planar limit: on the one hand, they can
be extrapolated to $\halfDp = 0$, leading to an estimate for the
planar Tolman length and the curvature dependence of $\surfacetension$
to first order in terms of $\halfDp$ or $1\slash\laplace$; on the other
hand, the slope of $\eer$ is in itself relevant, since its
zero-curvature limit $\eer_1$ also provides information on $\eer_0$.

The equivalent of the exact Tolman equation
in terms of the excess equimolar radius $\eer$ and the pressure
difference characterized by $\halfDp$ is Eq.\ (\ref{eqn:tolmanetafull}).
An expansion as a power series, analogous to Eq.\
(\ref{eqn:block}), can be expressed as
\begin{equation}
   \surfacetension = \planartension + 2\eer_0\halfDp
      - \frac{2\eer_0^2}{\planartension}\halfDp^2 + \landau{\halfDp^3}.
\label{eqn:blocketa}
\end{equation}
The planar limit, where higher order terms can be neglected, can be treated
accurately with expressions like Eq.\ (\ref{eqn:blocketa}).
Away from the planar limit, Eq.\ (\ref{eqn:tolmanetafull})
applies without any further condition (since the boundary
condition for the slope of $\eer$ is only relevant for $\halfDp \to 0$),
while Eq.\ (\ref{eqn:blocketa}) becomes an approximation.

%

%

%

%

%
\begin{figure}[b!]
\centering
\includegraphics[width=8.667cm]{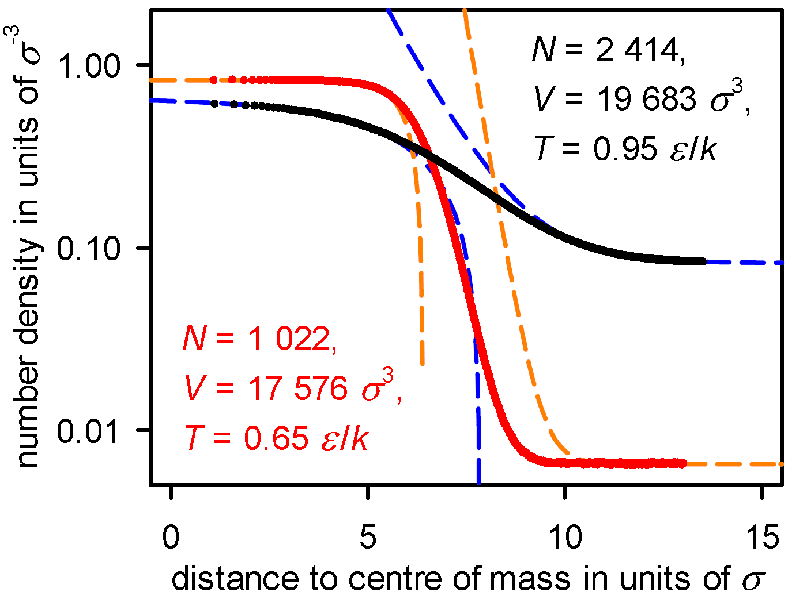}
\caption{
   Density profiles from canonical MD simulations of \LJTS{} liquid
   drops at $\temperature$ = $0.65$ and
   $0.95$ $\LJenergy\slash\kboltz$, showing the
   average densities from simulation ($\bullet$) and exponential
   approximants (-- --). The steeper profile corresponds to the lower
   temperature.
}
\label{fig:profile-temperature}
\end{figure}

\section{The excess equimolar radius from molecular simulation}
\label{sec:method}

With the \mardyn{} MD program, developed by Bernreuther and co-workers \cite{BV05, BNHVDHB09,
BBV10}, the canonical ensemble was simulated for small systems, corresponding
to equilibrium conditions for nanoscopic liquid drops surrounded by supersaturated
vapours. The truncated-shifted Lennard-Jones (\LJTS) pair potential
\begin{equation}
   \potential(\distance) = \left\{
      \begin{array}{ll}
         4\LJenergy\left[\left(\frac{\LJsize}{\distance}\right)^{12}
            - \left(\frac{\LJsize}{\distance}\right)^6\right] + \potential_\textnormal{shift},
            & \textnormal{\,\,for\,\,} \distance < \cutoff, \\
         0, & \textnormal{\,\,for\,\,} \distance \geq \cutoff, 
      \end{array}
   \right.
\end{equation}
with the size parameter $\LJsize$, the energy parameter $\LJenergy$ and
a cutoff at $\cutoff$ = $2.5$ $\LJsize$ is applied as a fluid model here,
including a shift by $\potential_\textnormal{shift}$ =
$4\LJenergy\left[(\LJsize\slash\cutoff)^6 - (\LJsize\slash\cutoff)^{12}\right]$
to make the potential continuous.
The \LJTS{} model is an adequate basis for investigating bulk and
interfacial properties of simple spherical conformal fluids
(e.g., noble gases and methane) on a molecular level,
cf.\ Vrabec \etal\ \cite{VKFH06}. On account of this,
numerous studies on nanoscopic liquid drops have been
reported \cite{LBYZ05, VKFH06, HL08, MPSF08, GB09, BDOVB10, HVH08, NJV10}.
The \LJTS{} fluid can thus be regarded as a key benchmark for theoretical
and simulation approaches to the problem of curved vapour-liquid interfaces.

%
\begin{figure}[h!]
\centering
\includegraphics[width=8.667cm]{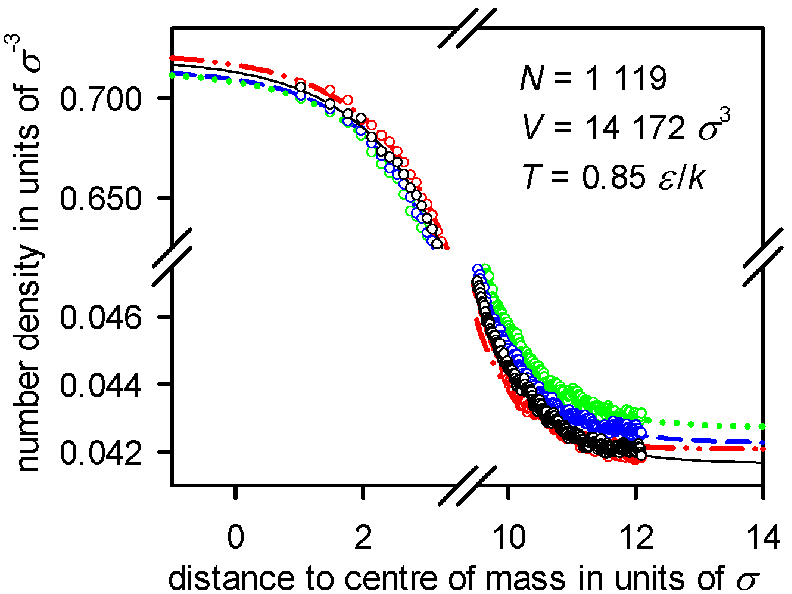}
\caption{
   Density profiles from a single canonical MD simulation of a \LJTS{}
   liquid drop at $\temperature$ = 0.85 $\LJenergy\slash\kboltz$, showing the
   average densities from simulation ($\circ$) and exponential
   approximants (lines) corresponding to the sampling intervals
   $2$ $000$ -- $3$ $000$ ($\cdots$; green), $3$ $000$ -- $4$ $000$ ($\cdots$ -- $\cdots$; red),
   $4$ $000$ -- $5$ $000$ (-- --; blue) and $5$ $000$ -- $6$ $000$ time units (---; black)
   after the onset of the simulation. The standard deviation between the
   densities at an infinite distance from the interface, according to
   the exponential fits for all sampling intervals of a single MD
   simulation, is used to determine the error of the bulk
   densities here.
}
\label{fig:profile-time}
\end{figure}

Certain of the general properties of this simple model, taking only
short-range interactions into account, can be assumed to carry over to polar fluids
as well \cite{Nezbeda05}, except for temperatures in the vicinity of the critical point.
It is clear, however, that a qualitatively different behaviour should be expected
for liquid drops formed by water with and without ionic species \cite{HA08, Galamba10},
liquid crystals \cite{HRR09} and similar complex organic molecules.
Such systems are beyond the scope of the present study.

Liquid drops are investigated at temperatures between
$\temperature$ = $0.65$ and $0.95$ $\LJenergy\slash\kboltz$, covering most of the
range between the triple point temperature (which is $\approx 0.55$
according to Bolhuis and Chandler \cite{BC00}, $\approx$ $0.618$ as determined by Toxv\ae{}rd
\cite{Toxvaerd07} and $\approx$ $0.65$ $\LJenergy\slash\kboltz$ according to van Meel \etal\ \cite{MPSF08})
and the critical temperature which several independent studies have consistently obtained as $1.08$
$\LJenergy\slash\kboltz$ for the \LJTS{} fluid \cite{SJ01, Smit02, VKFH06}.
The Verlet leapfrog algorithm is employed to solve the classical equations of
motion numerically with an integration time step of $0.002$ in Lennard-Jones
time units, i.e., $\LJsize\sqrt{\mass\slash\LJenergy}$, where $\mass$ is the
mass of a particle.
Cubic simulation volumes with $290$ to $126$ $000$ particles, applying the periodic
boundary condition, are equilibrated for at least $2$ $000$ time units.
Subsequently, spherically averaged density profiles $\density(\polar)$, with
their origin ($\polar = 0$) at the centre of mass of the whole system, are
constructed with a binning scheme based on equal volume concentrical spheres
using sampling intervals between $1\ 000$ and $40\ 000$ time units, depending
on the (expected) total simulation time, to gather multiple
samples for each system. Examples of the density profiles obtained
according to this method are shown in Figs.\ \ref{fig:profiles075} -- \ref{fig:profile-time}.

%
\begin{table*}[p!]
\caption{
   \label{tab:error}
   An analysis of the error of the excess equimolar radius $\eer$ of \LJTS{}
   liquid drops at the temperature $\temperature$ = $0.75$ $\LJenergy\slash\kboltz$.
   The number of particles $\absnum$, the volume $\volume$ of the periodic
   simulation box and the total simulation time $\simtime$ for the
   simulations of the liquid drops are indicated alongside
   the contributions to the uncertainty of $\eer$ from
   the pressure $\liq\pressure$ of the liquid phase
   (determined by canonical MD simulation of the bulk liquid),
   the surface tension $\planartension$ of the planar vapour-liquid interface, cf.\
   Vrabec et al.\ \cite{VKFH06}, the vapour pressure $\vap\pressure$ (analogous
   to $\liq\pressure$) and the equimolar radius $\equimolar$ (from the density
   profiles of the liquid drops). Note that the time
   unit, i.e., $\LJsize\sqrt{\mass\slash\LJenergy}$, corresponds to $500$ simulation
   time steps here.
   The flat symbols ($\flat$) indicate the
   fraction of the margin of error for $\eer$ due to the respective quantities.
   All values are given in Lennard-Jones units, and the
   error in terms of the last digit is specified in parentheses.
   In the subsequent discussion, the cases where the uncertainty of $\eer$
   exceeds $\LJsize$ are disregarded.
}
\begin{ruledtabular}
\begin{tabular}{ccc||cc|c|cc|cc||c}
$\absnum$ & $\volume$ [$\LJsize^{-3}$] & $\simtime$ [$\LJsize\sqrt{\mass\slash\LJenergy}$]
   & $\liq\pressure$ [$\LJenergy\LJsize^{-3}$] & $\contribution{\liq\pressure}$ & $\contribution{\planartension}$
      & $\vap\pressure$ [$\LJenergy\LJsize^{-3}$] & $\contribution{\vap\pressure}$
         & $\equimolar$ [$\LJsize$] & $\contribution{\equimolar}$ & $\eer$ [$\LJsize$] \\ \hline

$\phantom{000}$ $497$ & $\phantom{0}10$ $648$ & $\phantom{0}60$ $000$ 
   & $0.6(1)\phantom{00}$ & $84 \,\%$ & $\phantom{0}5.9 \,\%$
      & $0.0135(3)\phantom{0}$ & $0.23 \,\%$
         & $\phantom{0}4.33(5)\phantom{0}$ & $9.5 \,\%\phantom{0}$ & $\phantom{-}2.5(5)$ \\
$\phantom{00}1$ $418$ & $\phantom{0}21$ $952$ & $\phantom{0}48$ $176$ 
   & $0.16(1)\phantom{0}$ & $82 \,\%$ & $17 \,\%\phantom{.0}$
      & $0.01136(5)$ & $0.37 \,\%$
         & $\phantom{0}6.883(3)$ & $0.5 \,\%\phantom{0}$ & $\phantom{-}0.4(6)$ \\
$\phantom{00}1$ $766$ & $\phantom{0}21$ $952$ & $\phantom{00}6$ $000$ 
   & $0.14(3)\phantom{0}$ & $94 \,\%$ & $\phantom{0}5.3 \,\%$
      & $0.0110(2)\phantom{0}$ & $0.58 \,\%$
         & $\phantom{0}7.61(1)\phantom{0}$ & $0.55 \,\%$ & $\phantom{-}0(2)\phantom{.0}$ \\
$\phantom{00}3$ $762$ & $\phantom{0}39$ $304$ & $221$ $244$ 
   & $0.113(2)$ & $53 \,\%$ & $45 \,\%\phantom{.0}$
      & $0.01042(4)$ & $1.0 \,\%\phantom{0}$
         & $\phantom{0}9.977(1)$ & $0.28 \,\%$ & $\phantom{-}0.3(3)$ \\
$\phantom{00}5$ $161$ & $\phantom{0}54$ $872$ & $\phantom{0}64$ $219$ 
   & $0.096(3)$ & $63 \,\%$ & $34 \,\%\phantom{.0}$
      & $0.0104(1)\phantom{0}$ & $2.4 \,\%\phantom{0}$
         & $11.089(4)$ & $0.82 \,\%$ & $-0.4(5)$ \\
$\phantom{00}6$ $619$ & $\phantom{0}74$ $088$ & $162$ $678$ 
   & $0.090(2)$ & $59 \,\%$ & $40 \,\%\phantom{.0}$
      & $0.01007(2)$ & $0.75 \,\%$
         & $12.029(3)$ & $0.57 \,\%$ & $-0.2(5)$ \\
$\phantom{0}10$ $241$ & $110$ $592$ & $185$ $460$ 
   & $0.080(1)$ & $56 \,\%$ & $43 \,\%\phantom{.0}$
      & $0.00985(2)$ & $0.58 \,\%$
         & $13.974(2)$ & $0.29 \,\%$ & $-0.1(5)$ \\
$\phantom{0}12$ $651$ & $140$ $608$ & $\phantom{0}32$ $594$ 
   & $0.075(2)$ & $66 \,\%$ & $32 \,\%\phantom{.0}$
      & $0.00974(4)$ & $1.2 \,\%\phantom{0}$
         & $14.981(6)$ & $0.78 \,\%$ & $-0.2(8)$ \\
$\phantom{0}15$ $237$ & $166$ $375$ & $135$ $348$ 
   & $0.070(2)$ & $66 \,\%$ & $33 \,\%\phantom{.0}$
      & $0.00969(1)$ & $0.35 \,\%$
         & $15.967(1)$ & $0.15 \,\%$ & $-0.5(8)$ \\
$\phantom{0}17$ $113$ & $169$ $418$ & $\phantom{00}6$ $006$ 
   & $0.08(1)\phantom{0}$ & $89 \,\%$ & $\phantom{0}9.8 \,\%$
      & $0.00969(9)$ & $0.78 \,\%$
         & $16.689(4)$ & $0.18 \,\%$ & $\phantom{-}2(2)\phantom{.0}$ \\
$\phantom{0}24$ $886$ & $238$ $328$ & $\phantom{0}27$ $272$ 
   & $0.069(9)$ & $90 \,\%$ & $\phantom{0}9.7 \,\%$
      & $0.00947(3)$ & $0.3 \,\%\phantom{0}$
         & $18.969(5)$ & $0.17 \,\%$ & $\phantom{-}2(3)\phantom{.0}$ \\
$\phantom{0}28$ $327$ & $238$ $328$ & $\phantom{00}6$ $006$ 
   & $0.056(9)$ & $92 \,\%$ & $\phantom{0}7.5 \,\%$
      & $0.00945(3)$ & $0.28 \,\%$
         & $19.950(8)$ & $0.18 \,\%$ & $-1(5)\phantom{.0}$ \\
$\phantom{0}38$ $753$ & $247$ $673$ & $\phantom{00}6$ $000$ 
   & $0.050(7)$ & $90 \,\%$ & $\phantom{0}8.8 \,\%$
      & $0.00932(5)$ & $0.69 \,\%$
         & $22.391(7)$ & $0.15 \,\%$ & $-2(4)\phantom{.0}$ \\
$125$ $552$ & $697$ $078$ & $\phantom{00}6$ $006$ 
   & $0.042(5)$ & $89 \,\%$ &  $\phantom{0}9.6 \,\%$
      & $0.00908(9)$ & $1.6 \,\%\phantom{0}$
         & $33.31(1)\phantom{0}$ & $0.21 \,\%$ & $\phantom{-}3(5)\phantom{.0}$
\end{tabular}
\end{ruledtabular}
\end{table*}

%
\begin{table*}[p!]
\caption{
   \label{tab:presentsim}
   Number of particles $\absnum$, volume $\volume$ of the
   periodic simulation box and temperature $\temperature$ of the present canonical
   ensemble MD simulations of the \LJTS{} fluid and equilibrium properties
   of the liquid drop as well as the surrounding vapour, i.e., the respective
   densities $\liq\density,$ $\vap\density$ and pressures $\liq\pressure,$
   $\vap\pressure$ as well as the capillarity radius $\capillarity$, the
   equimolar radius $\equimolar$ and the excess equimolar radius $\eer$.
   For drop radii above $8$ $\LJsize$, these values can be reliably regarded
   as identical with those corresponding to the present theoretical approach, which
   is highlighted with the bold typeface. In case of smaller radii (cursive
   typeface), inaccuracies can arise due to
   the application of exponential approximants, cf.\ Fig.\ \ref{fig:profile-temperature}
   and Eq.\ (\ref{eqn:approximant}),
   so that the respective values can, at present, be acknowledged
   as phenomenological quantities only.
   All values are given in Lennard-Jones units, and the
   error in terms of the last digit is specified in parentheses.
}
\begin{ruledtabular}
\begin{tabular}{ccc||cc|cc||cc|c} 
   $\absnum$ & $\volume$ [$\LJsize^3$] & $\temperature$ [$\LJenergy\slash\kboltz$]
             & $\liq\density$ [$\LJsize^{-3}$] & $\vap\density$ [$\LJsize^{-3}$]
             & $\liq\pressure$ [$\LJenergy\LJsize^{-3}$] & $\vap\pressure$ [$\LJenergy\LJsize^{-3}$]
             & $\capillarity$ [$\LJsize$] & $\equimolar$ [$\LJsize$]
             & $\eer$ [$\LJsize$] \\ \colrule

   $\phantom{00}$ $291$ & $\phantom{00}8$ $999$ & $0.65$ 
              & $\mathit{0.857(5)\phantom{0}}$ & $\mathit{0.0090(2)\phantom{0}}$ & $\mathit{0.65(8)\phantom{00}}$ & $\mathit{0.0054(1)\phantom{0}}$
              & $\mathit{\phantom{0}2.1(3)}$ & $\mathit{\phantom{0}3.90(1)\phantom{0}}$ & $\mathit{\phantom{-}1.8(3)}$ \\
   $\phantom{0}1$ $022$ & $\phantom{0}17$ $576$ & $0.65$ 
              & $\mathit{0.830(1)\phantom{0}}$ & $\mathit{0.00651(7)}$ & $\mathit{0.22(2)\phantom{00}}$ & $\mathit{0.00397(4)}$
              & $\mathit{\phantom{0}6.3(6)}$ & $\mathit{\phantom{0}6.407(2)}$ & $\mathit{\phantom{-}0.1(6)}$ \\ \colrule

   $\phantom{00}$ $497$ & $\phantom{0}10$ $648$ & $0.75$ 
              & $\mathit{0.81(1)\phantom{00}}$ & $\mathit{0.0214(6)\phantom{0}}$ & $\mathit{0.6(1)\phantom{000}}$ & $\mathit{0.0135(4)\phantom{0}}$
              & $\mathit{\phantom{0}1.8(5)}$ & $\mathit{\phantom{0}4.33(5)\phantom{0}}$ & $\mathit{\phantom{-}2.5(5)}$ \\
   $\phantom{0}1$ $418$ & $\phantom{0}21$ $952$ & $0.75$ 
              & $\mathit{0.777(1)\phantom{0}}$ & $\mathit{0.0173(1)\phantom{0}}$ & $\mathit{0.16(1)\phantom{00}}$ & $\mathit{0.01136(5)}$
              & $\mathit{\phantom{0}6.5(6)}$ & $\mathit{\phantom{0}6.883(3)}$ & $\mathit{\phantom{-}0.4(6)}$ \\
   $\phantom{0}3$ $762$ & $\phantom{0}39$ $304$ & $0.75$ 
              & $\mathbf{0.7721(2)}$ & $\mathbf{0.01566(6)}$ & $\mathbf{0.113(2)\phantom{0}}$ & $\mathbf{0.01042(4)}$
              & $\mathbf{\phantom{0}9.7(4)}$ & $\mathbf{\phantom{0}9.977(1)}$ & $\mathbf{\phantom{-}0.3(4)}$ \\
   $\phantom{0}5$ $161$ & $\phantom{0}54$ $872$ & $0.75$ 
              & $\mathbf{0.7703(2)}$ & $\mathbf{0.0156(2)\phantom{0}}$ & $\mathbf{0.096(3)\phantom{0}}$ & $\mathbf{0.0104(1)\phantom{0}}$
              & $\mathbf{11.5(5)}$ & $\mathbf{11.089(4)}$ & $\mathbf{-0.5(6)}$ \\
   $\phantom{0}6$ $619$ & $\phantom{0}74$ $088$ & $0.75$ 
              & $\mathbf{0.7697(2)}$ & $\mathbf{0.01506(4)}$ & $\mathbf{0.091(2)\phantom{0}}$ & $\mathbf{0.01007(2)}$
              & $\mathbf{12.3(5)}$ & $\mathbf{12.029(3)}$ & $\mathbf{-0.2(5)}$ \\
   $10$ $241$ & $110$ $592$ & $0.75$ 
              & $\mathbf{0.7685(1)}$ & $\mathbf{0.01469(3)}$ & $\mathbf{0.080(2)\phantom{0}}$ & $\mathbf{0.00985(2)}$
              & $\mathbf{14.1(5)}$ & $\mathbf{13.974(2)}$ & $\mathbf{-0.1(5)}$ \\
   $12$ $651$ & $140$ $608$ & $0.75$ 
              & $\mathbf{0.7679(2)}$ & $\mathbf{0.01451(7)}$ & $\mathbf{0.075(2)\phantom{0}}$ & $\mathbf{0.00974(4)}$
              & $\mathbf{15.2(8)}$ & $\mathbf{14.981(6)}$ & $\mathbf{-0.2(8)}$ \\
   $15$ $237$ & $166$ $375$ & $0.75$ 
              & $\mathbf{0.7673(2)}$ & $\mathbf{0.01442(2)}$ & $\mathbf{0.070(2)\phantom{0}}$ & $\mathbf{0.00969(1)}$
              & $\mathbf{16.5(8)}$ & $\mathbf{15.967(1)}$ & $\mathbf{-0.5(8)}$ \\ \colrule

   $\phantom{0}1$ $119$ & $\phantom{0}14$ $172$ & $0.85$ 
              & $\mathit{0.733(7)\phantom{0}}$ & $\mathit{0.0421(5)\phantom{0}}$ & $\mathit{0.23(5)\phantom{00}}$ & $\mathit{0.0273(2)\phantom{0}}$
              & $\mathit{\phantom{0}3.1(9)}$ & $\mathit{\phantom{0}6.79(6)\phantom{0}}$ & $\mathit{\phantom{-}2.5(9)}$ \\
   $\phantom{0}3$ $357$ & $\phantom{0}32$ $768$ & $0.85$ 
              & $\mathbf{0.7135(8)}$ & $\mathbf{0.0371(5)\phantom{0}}$ & $\mathbf{0.097(5)\phantom{0}}$ & $\mathbf{0.0249(2)\phantom{0}}$
              & $\mathbf{\phantom{0}8.8(8)}$ & $\mathbf{\phantom{0}9.11(1)\phantom{0}}$ & $\mathbf{\phantom{-}0.4(9)}$ \\ \colrule

   $\phantom{0}2$ $031$ & $\phantom{0}21$ $952$ & $0.9\phantom{0}$ 
              & $\mathit{0.687(3)\phantom{0}}$ & $\mathit{0.0573(8)\phantom{0}}$ & $\mathit{0.13(1)\phantom{00}}$ & $\mathit{0.0369(3)\phantom{0}}$
              & $\mathit{\phantom{0}5.1(8)}$ & $\mathit{\phantom{0}6.79(6)\phantom{0}}$ & $\mathit{\phantom{-}1.7(9)}$ \\
   $\phantom{0}4$ $273$ & $\phantom{0}29$ $791$ & $0.9\phantom{0}$ 
              & $\mathbf{0.6773(9)}$ & $\mathbf{0.0532(2)\phantom{0}}$ & $\mathbf{0.082(4)\phantom{0}}$ & $\mathbf{0.03516(7)}$
              & $\mathbf{\phantom{0}9.7(9)}$ & $\mathbf{10.086(9)}$ & $\mathbf{\phantom{-}0.4(9)}$ \\
   $11$ $548$ & $\phantom{0}85$ $184$ & $0.9\phantom{0}$ 
              & $\mathbf{0.6738(1)}$ & $\mathbf{0.0504(2)\phantom{0}}$ & $\mathbf{0.0672(6)}$ & $\mathbf{0.03396(8)}$
              & $\mathbf{13.7(4)}$ & $\mathbf{14.054(8)}$ & $\mathbf{\phantom{-}0.4(4)}$ \\ \colrule

   $\phantom{0}2$ $414$ & $\phantom{0}19$ $683$ & $0.95$ 
              & $\mathit{0.662(2)\phantom{0}}$ & $\mathit{0.0825(2)\phantom{0}}$ & $\mathit{0.169(7)\phantom{0}}$ & $\mathit{0.05032(8)}$
              & $\mathit{\phantom{0}2.7(3)}$ & $\mathit{\phantom{0}6.86(3)\phantom{0}}$ & $\mathit{\phantom{-}4.2(3)}$ \\ 
\end{tabular}
\end{ruledtabular}
\end{table*}

The density profiles of \LJTS{} vapour-liquid interfaces
are known to agree well with an expression based
on two hyperbolic tangent terms, to which $\density(\polar)$ has been successfully
correlated for liquid drops by Vrabec \etal\ \cite{VKFH06}.
The present method, however, merely requires the bulk
densities $\liq\density$ and $\vap\density$ corresponding
to a certain value of $\chempot$ or $\halfDp$, which are determined here by correlating
the outer part of the density profile and extrapolating it to
regions far from the interface. The densities of the coexisting fluid phases
are thus deduced from simulation results by correlating the exponential terms
\begin{eqnarray}
   \liq\density & = & \density(\polar)
      + \liq\parama\exp\left(\liq\paramb[\polar - \liq\polar]\right), \nonumber\\
   \vap\density & = & \density(\polar)
      - \vap\parama\exp\left(\vap\paramb[\vap\polar - \polar]\right),
\label{eqn:approximant}
\end{eqnarray}
to the data for the inner- and outermost spherical bins of the density profiles,
cf.\ Fig.\ \ref{fig:profile-temperature}.
These terms, which are based on those employed by Lekner and Hender\-son \cite{LH77},
asymptotically agree with the hyperbolic tangent expression of Vrabec \etal\ \cite{VKFH06}.
From the liquid and vapour densities $\liq\density$ and $\vap\density$ of
the fit to Eq.\ (\ref{eqn:approximant}), the equimolar radius $\equimolar$ is calculated
according to Eq.\ (\ref{eqn:equimolar}). The respective margins of error are
obtained as standard deviations from the profiles belonging to different
sampling intervals of the same MD simulation, cf.\ Fig.\ \ref{fig:profile-time},
of which there are at least three in all cases.
The corresponding pressures $\liq\pressure$ and $\vap\pressure$ are computed by
canon\-ical MD simulation of the bulk fluid at the respective densities.

For the surface tension in the zero-curvature limit, the values
$\planartension(0.65 \, \LJenergy\slash\kboltz) = 0.680 \pm 0.009,$
$\planartension(0.75 \, \LJenergy\slash\kboltz) = 0.493 \pm 0.008,$ 
$\planartension(0.85 \, \LJenergy\slash\kboltz) = 0.317 \pm 0.007$ and
$\planartension(0.95 \, \LJenergy\slash\kboltz) = 0.158 \pm 0.006$ $\LJenergy\LJsize^{-2}$
are taken from the correlation of Vrabec \etal\ \cite{VKFH06}; the
error corresponds to the individual data points for $\planartension$ from the same source.
In case of $\temperature = 0.9$ $\LJenergy\slash\kboltz,$ the higher
precision of the computations of van Giessen and Blokhuis \cite{GB09} is exploited,
using the value $\planartension = 0.227 \pm 0.002$ $\LJenergy\LJsize^{-2}$ obtained from
a linear fit to data for the curved interface \cite{GB09}, cf.\ Fig.\ \ref{fig:VGB}.
The assumption made for the error is rather generous in this case, considering the
even higher confidence suggested by the agreement between the individual data
points for $\halfDp\equimolar$.

%
\begin{figure}[t!]
\centering
\includegraphics[width=8.667cm]{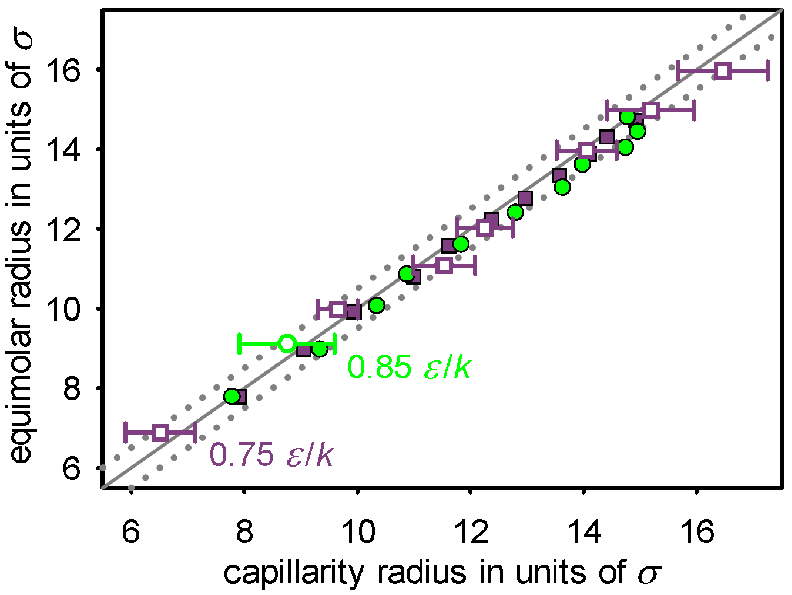}
\caption{
   \label{fig:P-Q}
   Equimolar radius $\equimolar$ as a function of the capillarity radius $\capillarity$
   for LJTS liquid drops, from density profiles and bulk pressures
   determined with canonical MD simulations at $\temperature$ = $0.75$ ($\square$)
   and $0.85$ $\LJenergy\slash\kboltz$ ($\circ$), in
   comparison with results from previous work of
   Vrabec \etal\ \cite{VKFH06} at $\temperature$ = $0.75$ ($\blacksquare$)
   and $0.85$ $\LJenergy\slash\kboltz$ ($\bullet$),
   using pressure differences based on evaluating the IK tensor in the
   (approximately) homogeneous regions inside and outside the liquid
   drop. The continuous diagonal line is defined by $\equimolar = \capillarity$ and
   thus corresponds to an excess equimolar radius of $\eer = 0$, while the dotted
   lines correspond to $\eer = \pm 0.5$ $\LJsize$.
}
\end{figure}

Combining these quantities leads to the capillarity radius $\capillarity$ and the excess
equimolar radius $\eer$. Note that the margin of error for $\eer$, as indicated in
Tab.\ \ref{tab:error}, contains contributions quantifying the accuracy of $\planartension$ and the
precision of the MD simulations of the liquid drop itself as well as those of
the homogeneous vapour and liquid phases. While the vapour pressure $\vap\pressure$
and the equimolar radius $\equimolar$ could be obtained with a high precision,
the liquid pressure and the surface tension in the zero-curvature limit are 
major sources of uncertainty here. In both cases, methodical changes
can be expected to increase the precision significantly: regarding $\planartension$,
it can be seen from Fig.\ \ref{fig:VGB} that it is now possible to reach
a level of confidence beyond that of the data of Vrabec \etal\ \cite{VKFH06} which
are also used here.
For the pressure of the liquid $\liq\pressure$,
approaches based on the chemical potential, which can be
determined in any region of the simulation volume (including the vapour phase),
can be expected to lead to significant improvements
in combination with a reliable equation of state or high-precision simulations
in the grand canonical ensemble.

A full summary of the simulation results where $\eer$ could be determined with
error bars smaller than $\LJsize$ is given in Tab.\ \ref{tab:presentsim}.
Note that to achieve full consistency with the Tolman approach, the
bulk densities $\liq\density$ and $\vap\density$ from Eq.\ (\ref{eqn:approximant}) have to match
those of the bulk fluid at the same temperature and chemical potential
as the two-phase system. Regarding liquid drops with $\equimolar$ $>$ $8$ $\LJsize$,
this is certainly the case, since constant density regions coexisting with the
interface are actually present, cf.\ Fig.\ \ref{fig:profiles075}.
However, the values determined for the smallest drops here rely on the
validity of the correlation given by Eq.\ (\ref{eqn:approximant}) and can be considered valid 
only as far as this expression itself does not introduce any major deviations,
an assertion that remains open to further examination; a version of the present
method computing $\liq\pressure$ via $\chempot$ could resolve this issue.

\section{Discussion} 
\label{sec:discussion}


Previous authors have made qualitatively contradictory claims on the
magnitude of the Tolman length as well as its sign: Tolman himself expected
$\tolman$ to be positive and smaller than the length scale of the dispersive
interaction, a conjecture that Kirkwood and Buff \cite{KB49} affirmed
from a statistical mechanical point of view, based on a mechanical approach.
Subsequent studies, however, have also found $\tolman$ to be negligible or even equal
to zero \cite{BD98, LBYZ05, ZZMZ11}, positive and larger than $\LJsize$ \cite{VKFH06, HVH08},
negative with $- \LJsize < \tolman < 0$ \cite{HHR81, GB09, SMMMJ10}
or negative and diverging ($\tolman_0 = - \infty$) in the planar limit \cite{BRMD03},
while others have claimed that the sign of $\tolman$ is curvature dependent
itself \cite{KZS98, JNMV10}.
Thereby, they have only proven the mutual inconsistency
of their assumptions and methods, while nothing is truely known about $\tolman$ and
the dependence of the surface tension on curvature.

The new approach introduced in Section \ref{sec:eer} is strictly
based on axiomatic thermodynamics and relies on the fact that $\tolman_0 = -\eer_0$
holds in the planar limit. From the values for $\eer$
reported in bold face in Tab.\ \ref{tab:presentsim}, corresponding to
$\equimolar$ $>$ $8$ $\LJsize$, the excess equimolar radius for liquid
drops of the \LJTS{} fluid is unequivocally shown to be smaller in magnitude
than $\LJsize \slash 2$, while its remains unclear whether it is positive,
negative, of both signs (depending on the curvature) or equal to zero.
Since this means that at the present level of accuracy, no significant
dependence of $\surfacetension$ on the radius of the liquid drop could be
detected, the statement of Mareschal \etal\ \cite{MBL97} regarding
cylindrical interfaces also applies here: considering \qq{the large
fluctuations in the bulk liquid phase}, cf.\ the error analysis presented
in Tab.\ \ref{tab:error}, \qq{we tentatively conclude that the surface
tension is independent of the curvature of the liquid-vapor interface
or else that this dependence is very weak.}

The only view that can be definitely dismissed is that
of a large and positive Tolman length, previously held by some of the
present authors on the basis of results from the mechanical route to the surface
tension, employing the IK pressure tensor \cite{VKFH06, HVH08}.
As Fig.\ \ref{fig:P-Q} shows, the previous simulation results are actually
consistent with those from the present study if they are interpreted
in terms of the radii $\capillarity$ and $\equimolar$. Thereby, following
the approach of van Giessen and Blokhuis \cite{GB09}, only
the density profile and the pressure in the homogeneous regions inside
and outside the liquid drop are taken into account, whereas the normal
pressure along the interface is not considered at all.
Since the deviation between present and previous data disappears
in such a representation, the disagreement must be caused by the inadequacy
of the pressure-tensor (mechanical) route implemented by Thompson \etal\ \cite{TGWCR84},
as pointed out by Henderson \cite{Henderson86, Henderson99}.
Possible sources of error for this approach are outlined in Section \ref{sec:review}.
Nonetheless, more detailed methodological investigations are expedient
to determine which approximations are actually responsible for
major inaccuracies, and whether they can be corrected or
whether the pressure-tensor route to the surface
tension has to be discarded altogether.

\begin{acknowledgments}
The present work contributes to
the IMEMO project of the German Federal Ministry of
Education and Research (BMBF) and to the Collaborative Research Centre (SFB)
926 of the German Research Foundation (DFG).
It was conducted under the auspices of
the Boltzmann-Zuse Society of Compuational Molecular Engineering (BZS).
The position of M.\ T.\ Horsch at Imperial College London was
funded by a fellowship within the postdoc programme of the German
Academic Exchange Service (DAAD), and G.\ Jackson as well as E.\ A.\
M\"uller are grateful to the Engineering and Physical Sciences
Research Council (EPSRC) of the UK (grants GR/T17595, GR/N35991 and
EP/E016340), the Joint Research Equipment Initiative (GR/M94427), and
the Royal Society-Wolfson Foundation refurbishment scheme for
additional funding to the Molecular Systems Engineering Group. 
The computations were performed on the NEC Nehalem cluster
\textit{laki} at the High Performance Computing Center Stutt\-gart (HLRS)
with resources allocated according to the grant MMHBF.
At the HLRS, the authors would like to thank M.\ F.\ Bernreuther for
his support in general and for co-ordinating the MMHBF grant as well as the
development of the MD code \mardyn.
Furthermore, D.\ Reguera L\'opez and J.\ Wedekind (Barcelona),
F.\ R\"omer (London), M.\ Schrader (Mainz), Z.\ Lin,
S.\ K.\ Miroshnichenko, S.\ Olma, Z.\ Wei (Paderborn) and D.\ V.\
Tatyanenko (St.\ Petersburg) as well as S.\ Dietrich,
S.\ Grottel, C.\ Niethammer and G.\ Reina (Stuttgart)
are acknowledged for contributing to
various theoretical and practical issues
through helpful suggestions and their participation in relevant discussions
or by assisting at the debugging process.
\end{acknowledgments}

\end{document}